\documentclass[%
 reprint,
 amsmath,amssymb,
 aps,
eprint
prl]{revtex4-2}

\usepackage{graphicx}
\newcommand{\cev}[1]{\reflectbox{\ensuremath{\vec{\reflectbox{\ensuremath{#1}}}}}}
\usepackage{graphicx}
\usepackage{dcolumn}
\usepackage{bm}
\usepackage{enumitem}
\usepackage[colorlinks=true]{hyperref}

\begin{document}

\preprint{APS/123-QED}
\title{Quantum Bipolar Thermoelectricity}
\author{F. Antola$^1$}
\email[Corresponding author: ]{filippo.antola@sns.it}
\affiliation{%
NEST, Istituto Nanoscienze-CNR and Scuola Normale Superiore, Pisa, Italy
}

\author{G. De Simoni$^1$}
%\email{giorgio.desimoni@nano.cnr.it}
\affiliation{%
Istituto Nanoscienze – CNR, NEST-SNS, Piazza San Silvestro 12, Pisa, Italy
}
\author{F. Giazotto$^1$}
%\email{francesco.giazotto@sns.it}
\affiliation{%
Istituto Nanoscienze – CNR, NEST-SNS, Piazza San Silvestro 12, Pisa, Italy
}
\author{A. Braggio$^{1,2}$}
%\email{alessandro.braggio@nano.cnr.it}
\affiliation{%
Istituto Nanoscienze – CNR, NEST-SNS, Piazza San Silvestro 12, Pisa, Italy
}
\affiliation{Institute for Quantum Studies, Chapman University, Orange, CA, USA
}

\begin{abstract}
Thermoelectricity is generally understood as a classical effect emerging from energy-dependent transport asymmetries. Here, we uncover a purely \emph{quantum} mechanism, where a superconducting S-I-S' tunnel junction in thermal equilibrium develops a nonlinear bipolar thermoelectric response owing to the dynamical Coulomb blockade and the emission-absorption imbalance of a cold electromagnetic bath. Two representative environments are analysed, revealing Seebeck coefficients up to $100~\mu\mathrm{V/K}$ for realistic junction parameters. Because the response directly reflects the spectral properties of the surrounding environment, our results suggest that bipolar quantum thermoelectricity could provide a new route for spectroscopic sensing of electromagnetic modes and for designing low-temperature thermoelectric devices with environmentally engineered performance.
\end{abstract}

\maketitle

\section*{Introduction}
A key question in quantum physics is how quantum fluctuations in the electromagnetic environment can induce complex electronic phenomena~\cite{devoret_effect_1990,fujisawa_spontaneous_1998,aguado_double_2000,henriet_electrical_2015}. Although thermoelectricity is usually understood within a classical diffusive framework~\cite{grosso_solid_2000}, it is natural to ask whether purely quantum features can also generate thermoelectric effects~\cite{benenti_fundamental_2017,whitney_most_2014}. This question has been investigated in quantum dots, where discrete energy levels enable energy filtering and thermoelectric conversion at the single-electron level~\cite{sanchez_optimal_2011,rossello_dynamical_2017,jordan_powerful_2013,thierschmann_three-terminal_2015,josefsson_quantum-dot_2018,jaliel_experimental_2019}.
Superconducting tunnel junctions, because of their sharp energy-dependent density of states (DOS), strongly enhance photon-assisted processes and are therefore highly sensitive to quantum fluctuations~\cite{deblock_detection_2003,billangeon_emission_2006,basset_emission_2010,basset_high-frequency_2012}. They have recently attracted attention as platforms for quantum thermal machines~\cite{pekola_normal-metal-superconductor_2007,hofer_quantum_2016,tan_quantum-circuit_2017,antola_tunable_2024}, as probes of electromagnetic environments~\cite{souquet_photon-assisted_2014,lorch_optimal_2018,mecklenburg_thermopower_2017,karimi_optimized_2020,nikolic_optimized_2023,cailleaux_theory_2025,hubler_nonclassical_2025} and for developing strong thermoelectric effects~\cite{machon_nonlocal_2013,ozaeta_predicted_2014,bergeret_colloquium_2018,marchegiani_nonlinear_2020,germanese_bipolar_2022}, which can in turn be exploited for the detection of radiation signals~\cite{heikkila_thermoelectric_2018,chakraborty_thermoelectric_2018,geng_superconductor-ferromagnet_2020}.

In this work, we introduce an entirely \emph{quantum} mechanism for generating bipolar thermoelectricity in a superconducting tunnel junction with asymmetric energy gaps, where the junction is kept in thermal equilibrium but coupled to a cold electromagnetic environment.
In quantum transport, linear thermoelectric generation emerges from the breaking of the energy symmetry of electronic states around the Fermi energy $E_F$~\cite{benenti_fundamental_2017,arrachea_thermoelectric_2025}. Here, we show that a thermoelectric response can arise even in an ideally energy-symmetric system, driven by an imbalance between photon emission and absorption at energy $\hbar\omega$, when coupled to a cold reservoir with $k_B T \lesssim \hbar\omega$.
Such a quantum regime can be realised in an S-I-S' tunnel junction between superconductors with different gaps $\Delta$ and $\Delta'$, where the relevant energy scale is $\hbar\omega\approx \Delta-\Delta'$~\cite{tinkham_introduction_2004,shapiro_superconductivity_1962,townsend_investigation_1962}. In these systems, the superconducting DOSs divergences enhance the weight of photon-assisted quasi-particle transitions, which dominate over the suppressed Josephson component\cite{marchegiani_phase-tunable_2020, germanese_phase_2023}. This suppression can be achieved through SQUID interferometry~\cite{germanese_bipolar_2022}, exploiting the Fraunhofer effect with an in-plane magnetic field~\cite{rowell_magnetic_1963}, or with very opaque tunnelling barriers with small Josephson energy~\cite{weides_barriers_2017}. 
\begin{figure*}[!t]
    \centering
    \includegraphics[width=\textwidth]{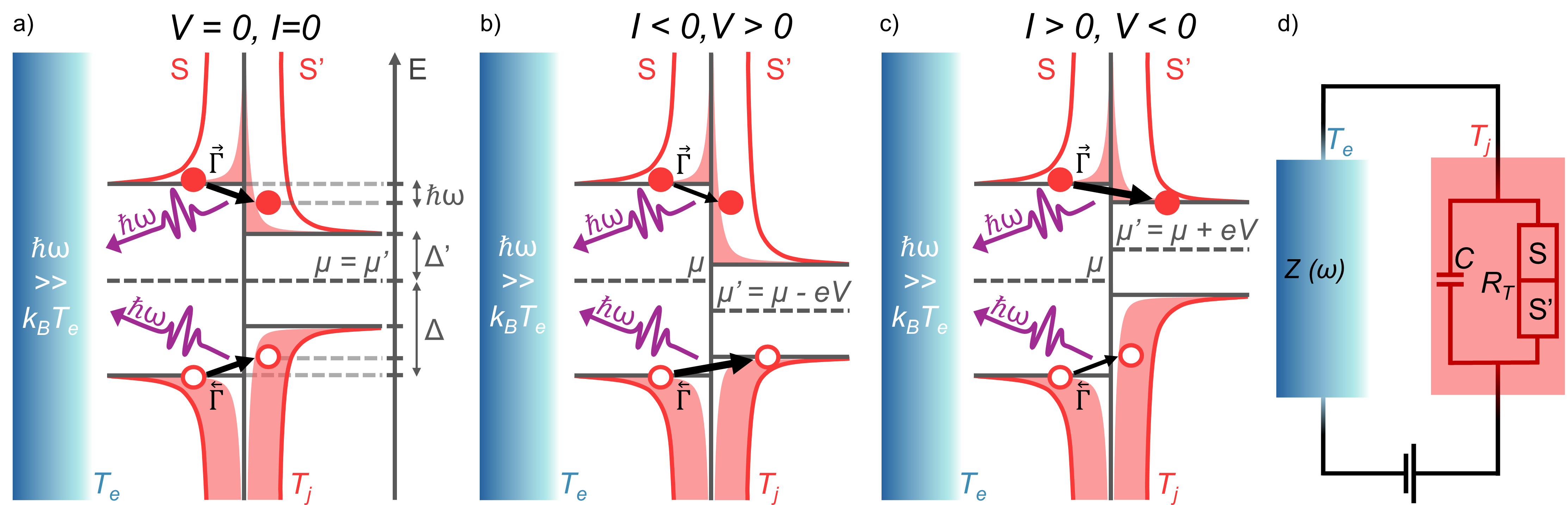}
    \caption{ \textbf{Quantum bipolar thermoelectricity from emission/absorption asymmetry}
    (a) Schematic of an S-I-S' junction at temperature $T_j$ coupled to a cold electromagnetic environment at temperature $T_e$, illustrating photon-assisted quasiparticle tunnelling processes at zero bias. In the quantum regime $k_B T_e \lesssim \hbar\omega$, the bath cannot emit $\hbar\omega$ photons, so only processes in which energy is transferred from the junction to the environment are allowed. Full (empty) red circles denote particle (hole) excitations. Hole-assisted transitions contribute to the backward rate, while particle-assisted transitions contribute to the forward rate. A vertical energy axis indicates the relevant quasiparticle energies. (b) Same schematic as in (a), but in the presence of a small positive bias $V>0$. (c) Same as in (b), but for a negative bias $V<0$, where the forward tunnelling rate becomes dominant. d)  Minimal circuit: an S-I-S' junction (capacitance $C$, tunnel resistance $R_T$) coupled to an environmental impedance $Z(\omega)$ and to an external bias.}
    \label{Fig 1}
\end{figure*}

Before explaining in detail the mechanism of quantum thermoelectricity generation, we briefly summarise why this can emerge in such simple tunnelling systems. Figure~\ref{Fig 1}(a) schematically illustrates an asymmetric-gap S-I-S' junction at temperature $T_j$, coupled to an electromagnetic environment at temperature $T_e$. Energy exchange with the junction is dominated by photon-assisted processes involving quanta of energy $\hbar\omega$, as determined by the energy selectivity of the S-I-S' junction superconducting DOSs. We focus on the quantum regime $k_B T_e \lesssim \hbar\omega$, where the environment is sufficiently cold to suppress photon emission from the bath. As a result, tunnelling occurs \emph{only} via processes in which energy quanta are transferred from the junction to the electromagnetic environment, and the schematic highlights these energy-emission processes. For these emission processes, hole-assisted transitions contribute to the backward rate $\cev{\Gamma}$, while particle-assisted transitions contribute to the forward rate $\vec{\Gamma}$, with dominant contributions arising from quasiparticle states near the superconducting gap edges (shown with arrows). For clarity, Fig.~\ref{Fig 1}(a) also includes a vertical energy axis indicating the relative energies of the relevant quasiparticle states.
In the following, we parametrize the applied bias by the voltage $V$, so that the difference in electrochemical potential between the electrodes is $\mu-\mu' = eV$. In the absence of bias ($eV=0$), the forward and backward tunnelling rates coincide [panel Fig.~\ref{Fig 1}(a)]. Since the electric current is given by 
\begin{align}\label{Eq. Current}
I(V)=e\,[\vec{\Gamma}(eV)-\cev{\Gamma}(eV)],
\end{align}
this immediately implies that $I(0)=0$. For a small positive bias $V>0$ [panel Fig.~\ref{Fig 1}(b)], the forward and backward rates become unequal. Owing to the monotonically decreasing superconducting density of states, hole-assisted processes dominate, so that the backward rate exceeds the forward rate. 
According to Eq.~(\ref{Eq. Current}), this yields an electric current flowing against bias, $IV<0$. This behaviour is the hallmark of the thermoelectric response discussed here. A similar argument also applies to the opposite case $V<0$. 
Since the rates satisfy $\cev{\Gamma}(eV)=\vec{\Gamma}(-eV)$, reversing the bias interchanges the particle and hole processes. Consequently, for $V<0$, the forward rate becomes dominant, as illustrated in panel Fig.~\ref{Fig 1}(c), again producing a current flowing against the applied bias. This mechanism induces a bipolar thermoelectric response that closely mirrors the conventional bipolar effect~\cite{marchegiani_nonlinear_2020, marchegiani_superconducting_2020,germanese_bipolar_2022}. In strong contrast with this known bipolar thermoelectric scenario, in which the effect arises from a temperature difference across the junction, the present response instead arises from a purely \emph{quantum} asymmetry in the radiative coupling between the S-I-S' junction and the surrounding cold electromagnetic environment, enabled by the temperature difference between the two. Moreover, the environment strongly affects transport \emph{only} when the junction operates in the dynamical Coulomb blockade (DCB) regime, such as in the presence of a high-impedance environment \cite{devoret_effect_1990, ingold_charge_1992, nazarov_quantum_2009}

\section*{Results}
\begin{figure*}[!t]
    \centering
    \includegraphics[width=\textwidth]{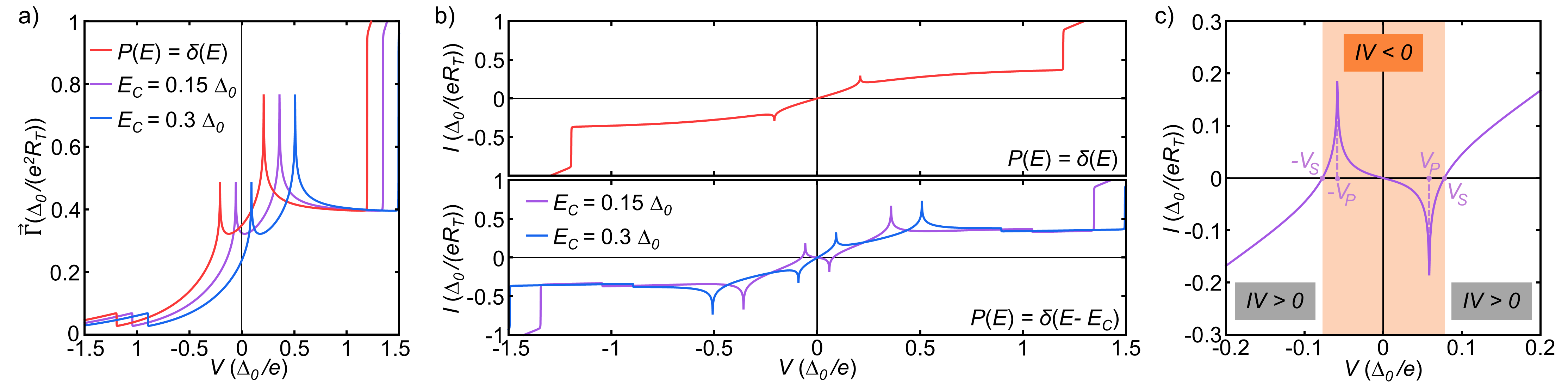}
    \caption{ \textbf{Quantum bipolar thermoelectricity and charging energy}
        a) Forward tunnelling rate $\vec{\Gamma}(eV)$ for two different charging energies $E_C$ in the high-impedance (violet/blue) and low-impedance (red) regimes. Other parameters: r=0.9 and $T_j = 0.8\ T_C$, where $T_C$ is the critical temperature of the S electrode. (b) Corresponding current-voltage ($I$-$V$) characteristics. c) Enlarged view of the $I$–$V$ curve for $E_C = 0.15\,\Delta_0$, showing the thermoelectric regime (orange area). The power-maximum voltage $V_P$ and the Seebeck voltage $V_S$ are marked.}
    \label{Fig 2}
\end{figure*}
\subsection*{Tunnelling Rates}
To quantitatively describe this situation, we consider the circuit shown in Fig.~\ref{Fig 1}(d), a superconducting tunnel junction of normal-state resistance $R_T$ and intrinsic capacitance $C$, coupled through dissipation-less wires to a remote external impedance $Z(\omega)$ that constitutes the electromagnetic environment. The total impedance seen by the junction is approximated as the parallel combination of its capacitance and the environmental impedance, $Z_t(\omega) = (i\omega C + Z^{-1}(\omega))^{-1}$.
For a junction in thermal equilibrium at $T_j$ and subject to a DC bias $\delta\mu=eV$, the forward tunnelling rate $\vec{\Gamma}(eV)$ reads~\cite{ingold_charge_1992}
\begin{align}
\label{Intro/rate fin}
\vec{\Gamma}(eV) =& \frac{1}{e^2 R_T} \int_{-\infty}^{\infty} dEdE'\ N_S(E) N_{S'}(E' + eV) \nonumber \\
& \times f(E)\left[1 - f(E' + eV)\right] P(E - E'),
\end{align}
where $f(E) = 1/(1 + e^{E/k_B T_j})$ is the Fermi-Dirac distribution. The functions $N_i(E)$, with $i=S,S'$, denote the normalized superconducting
densities of states of the electrodes, including the lifetime broadening effects accounted for by a finite Dynes parameter  $\gamma$~\cite{tinkham_introduction_2004}. The DOSs depend on the temperature-dependent superconducting gaps $\Delta=\Delta(T)$ of the two electrodes, whose zero-temperature values $\Delta_0$ and $\Delta_0^\prime$ define the device parameter $r=\Delta_0^\prime/\Delta_0$. A graphical representation highlighting the characteristic divergences of the superconducting densities of states is shown in Fig.~\ref{Fig 1}(a), while the explicit expressions used in the calculations are specified in the Methods. Fermi's golden rule rates contain a double integral due to energy exchange $\Delta E=E - E'$ with the electromagnetic environment. This exchange is described by the function $P(\Delta E)$, which gives the probability that a tunnelling event involves an energy transfer $\Delta E$ to or from the environment. The positive (negative) values of $\Delta E$ correspond to the energy emission (absorption) by the junction. The function is defined as $P(\Delta E) = \frac{1}{2\pi\hbar} \int_{-\infty}^{\infty} dt\ e^{J(t) + \frac{i}{\hbar} \Delta E t}$, where the function $J(t)$ describes the influence of the electromagnetic environment and explicitly depends on its impedance $Z_t(\omega)$ (see Methods Section). In particular, when the real part of the impedance is much smaller than the von Klitzing constant $R_K = h/e^2$, $\Re\left[Z_t(\omega)\right]\ll R_K$, the correlation function $J(t)$ vanishes and $P(E)\approx\delta(E)$, recovering the standard elastic tunnelling rate with a single-energy integral~\cite{tinkham_introduction_2004,marchegiani_nonlinear_2020}. In the opposite limit of a strongly dissipative environment, $\Re\left[Z_t(\omega)\right]\approx R \gg R_K$, inelastic processes dominate, leading to the physics of DCB~\cite{devoret_effect_1990,ingold_charge_1992,nazarov_quantum_2009}. \\
A similar expression to Eq.~(\ref{Intro/rate fin}) can also be written for the backward rate $\cev{\Gamma}(eV)$, and the tunnelling current can then be easily computed as given in Eq.~(\ref{Eq. Current}). When the system and the environment are \emph{all} at the same temperature, $T_j=T_e=T$, the tunnelling rates satisfy the detailed balance
$\vec{\Gamma}(-E) = e^{-\beta E} \vec{\Gamma}(E)$, with $\beta = 1/k_B T$. This relation is also reflected in the function $P(\Delta E)$, with $P(-\Delta E) = e^{-\beta \Delta E} P(\Delta E)$~\cite{ingold_charge_1992}. For symmetric DOS around the Fermi level, $N_i(E-E_F)=N_i(-(E-E_F))$, the current obeys the reciprocity relation $I(-V)=-I(V)$. In this regime of global thermal equilibrium, it reduces to $I(V)=e\vec{\Gamma}(eV)(1-e^{-\beta eV})$, implying that quasiparticle transport is always dissipative, i.e.,  $I(V)V > 0$, regardless of the nature of the electromagnetic environment and consistent with what has been reported before~\cite{marchegiani_nonlinear_2020}. 
Furthermore, the general reciprocity property ensures that $I(V{=}0)=0$, forbidding any linear thermoelectric effect. 
However, in what follows, we show that a nonlinear bipolar thermoelectric current of intrinsic \emph{quantum} nature can still be generated when the junction is in thermal equilibrium, provided that the electromagnetic environment is sufficiently cold. In this regime, the tunnelling current can flow against the applied bias, leading to $I(V)V<0$ and therefore to net power generation.

\subsection*{Dissipative Environment}
We begin by considering the simplest case of a purely resistive electromagnetic environment, modelled as a frequency-independent impedance, $Z(\omega) = R$.
A natural scale for the strength of the electromagnetic environment at the junction is the dimensionless conductance parameter $g = R_K / R$. 

In the limit of low environmental impedance ($g\to\infty$), even at zero bath temperature ($T_e \to 0$), there are no environmental corrections and $P(\Delta E)=\delta(\Delta E)$, recovering the standard elastic S-I-S' tunnelling process also called the "semiconductor" model~\cite{tinkham_introduction_2004}.
In this case, the tunnelling rate corresponds to the red line in Fig.~\ref{Fig 2}(a) and is independent of capacitance $C$. For $eV>0$, the rate increases sharply when the energy difference exceeds $eV=\Delta + \Delta'$, leading to Ohmic-like behaviour at higher bias. Below this threshold, in the subgap region, quasiparticle transport is strongly suppressed. However, at a finite junction temperature, a pronounced peak appears at the matching energy $eV=\Delta-\Delta'$, resulting from the alignment of the singularities in the DOSs of the two superconducting electrodes. The height of this peak is directly related to the junction temperature $T_j$~\cite{marchegiani_nonlinear_2020}.
At the same time, a secondary peak emerges at $eV=-(\Delta-\Delta')$, consistent with the detailed balance. Practically, for $g\to\infty$ the electromagnetic environment does not affect the junction, even if it is at a different temperature.

In the opposite regime of high impedance ($g \to 0$), the total effective impedance reduces to $\Re[Z_t(\omega)] \approx (\pi/C) \delta(\omega)$, allowing an analytical evaluation of the $P(E)$ function as $P(\Delta E)=\delta(\Delta E-E_C)$, where $E_C = e^2 / (2C)$ is the charging energy. In the zero-temperature limit, the environment can absorb only this additional energy.
The tunnelling rate, see violet and blue lines in Fig.~\ref{Fig 2}(a), becomes capacitance-dependent and retains the same shape as in the low-impedance case, but, indeed, shifted by Coulomb energy $E_C$. 
This energy shift also implies that detailed balance is necessarily violated due to energy transfer to the cold environmental impedance, since the whole system is no longer at equilibrium. 

Furthermore, when $E_C \lesssim \Delta - \Delta'$, the system presents a regime characterised by a "strong" violation of detailed balance \cite{battisti_bipolar_2024}, where $\vec{\Gamma}(-eV) > \vec{\Gamma}(eV)$. This phenomenon is clearly illustrated in Fig.~\ref{Fig 2}(a) (violet line), signalling the emergence of the bipolar thermoelectric effect induced by coupling with the cold electromagnetic environment.

The current, shown in Fig.~\ref{Fig 2}(b), confirms the scenario discussed. In the low impedance limit ($g\to \infty $, top panel), the $I$-$V$ characteristic is \emph{only} dissipative, as expected for an S-I-S' junction at equilibrium temperature $T_j$. In contrast, for $g\to 0$, the system is sensitive to the cold electromagnetic environment, which effectively acts as a cold reservoir. For suitable values of the charging energy $E_C$, a negative current peak emerges at $eV_P=\Delta-\Delta'-E_C$ when the "strong" violation of the detailed balance condition occurs in the rate. As highlighted by the orange-shaded region in Fig.~\ref{Fig 2}(c), the system exhibits a regime with negative absolute conductance, $I(V)V<0$, corresponding to net power generation rather than dissipation. In particular, the response remains bipolar, satisfying the antisymmetry relation $I(-V)=-I(V)$. This behaviour constitutes a bipolar thermoelectric effect, which emerges even though the S–I–S' junction remains locally in thermal equilibrium. A non-local temperature difference between the junction and the colder environmental impedance is \emph{necessary}, demonstrating the thermoelectric nature of this mechanism. A key distinction from the bipolar thermoelectric effect driven by a temperature difference across the junction~\cite{marchegiani_nonlinear_2020,marchegiani_superconducting_2020,germanese_bipolar_2022} lies in the position of the matching peak, which is shifted toward zero by $e^2/2C$ due to the DCB Coulomb gap. Similar charging-induced shifts were reported for the bipolar thermoelectric response of a single-electron transistor~\cite{battisti_bipolar_2024}, although in that case the two leads are \emph{not} in thermal equilibrium.
It is important to clarify that the discussed effect does not originate from a non-equilibrium ratchet mechanism~\cite{linke_experimental_1999,reichhardt_ratchet_2017,custer_ratcheting_2020}. Indeed, increasing the asymmetry by taking the smaller gap $\Delta'\to0$ causes the effect to completely disappear rather than improve. Moreover, the $I$-$V$ characteristic remains fully reciprocal, excluding the simple diode effect mechanism.
Finally, in our analysis, we neglect any heating of the environmental resistor due to coupling to the hotter junction, assuming that it remains well thermalised with the cryostat base temperature.

\begin{figure}
    \centering
   \includegraphics[width=\linewidth]{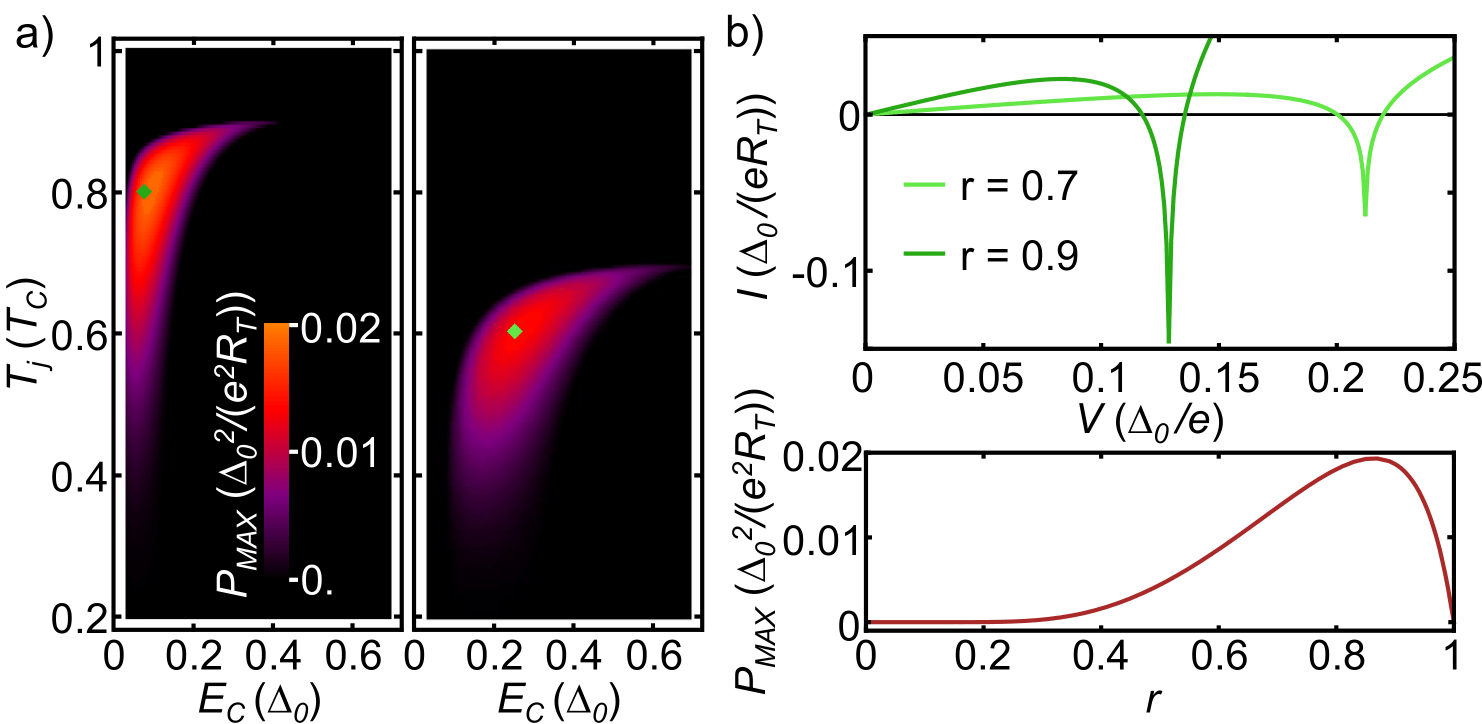}
    \caption{\textbf{Quantum thermoelectric performance} (a) Thermoelectric power at the matching peak $P_{MAX}$ as a function of $T_j$ and $E_C$, for $r = 0.9$ (left) and $r = 0.7$ (right), in the limit $T_e\to0$. Green diamond symbols mark the values of $E_C$ and $T_j$ that maximize the extracted power at the peak. (b) Top panel: $I$-$V$ characteristic at these same optimal conditions for $r=0.9$ and $r=0.7$. Bottom panel: $P_{MAX}$ computed at optimal $E_C$ and $T_j$ for different $r$.}
    \label{Fig 3}
\end{figure}

\subsection*{Quantum Thermoelectric Behaviour} In 
Fig.~\ref{Fig 3}(a) we show the maximal thermoelectric power, which is well approximated by the power value at the thermoelectric peak $V_P$, that is, $P_{\mathrm{MAX}} \approx |I(V_P) \cdot V_P|$, as a function of $E_C$ and $T_j$, for two values of the gap ratio $r$. The power becomes appreciable only above a specific temperature $T_j$, as a sufficient number of quasiparticle/quasihole states in the small gap side is necessary to probe the emission/absorption asymmetry.
The effect strengthens with increasing $T_j$ and reaches optimal performance around $T_j \approx 0.9 T_C'$, where $T_C'$ is the critical temperature of the lower-gap electrode $S'$. 
In the results shown, the temperature dependence of the superconducting gaps causes both superconducting gaps to decrease with increasing temperature (See Methods). However, their difference $\Delta(T_j) - \Delta'(T_j)$ increases, allowing the emergence of quantum thermoelectricity at a higher level $E_C$, as shown in the picture. For $\Delta_0 \simeq 0.18$~meV (Al) and $\Delta_0 \simeq 1.5$~meV (Nb), the maximum of the power color scale in Fig.~\ref{Fig 3}(a) corresponds to absolute values of order $\sim 5$~fW and $\sim 0.5$~pW, respectively, assuming $R_T=100~\mathrm{k}\Omega$.
The upper panel of Fig.~\ref{Fig 3}(b) shows the $I$-$V$ characteristics evaluated at the optimal operating points marked by the green diamonds in Fig.~\ref{Fig 3}(a), i.e., at the values of $(E_C,T_j)$ that maximize the extracted power at the matching peak. Under these conditions, the system exhibits a positive differential conductance near zero bias. This shows that, under certain conditions, the quantum bipolar thermoelectric effect is present only under strong nonlinear bias.
The lower panel of Fig.~\ref{Fig 3}(b) shows the extractable power as a function of the gap ratio $r$, evaluated at the values of $T_j$ and $E_C$ that maximize this quantity.
As $r$ increases, the power increases, peaks near $r \approx 0.9$, and then drops as the system approaches a symmetric S-I-S configuration. The disappearance of thermoelectricity can also be interpreted as a transition to the classical limit, where the symmetry between emission and absorption is again restored, since $(\Delta - \Delta')\approx\hbar\omega \sim k_B T_e$.

\subsection*{Quantum regime}
\begin{figure}
    \centering
    \includegraphics[width=\linewidth]{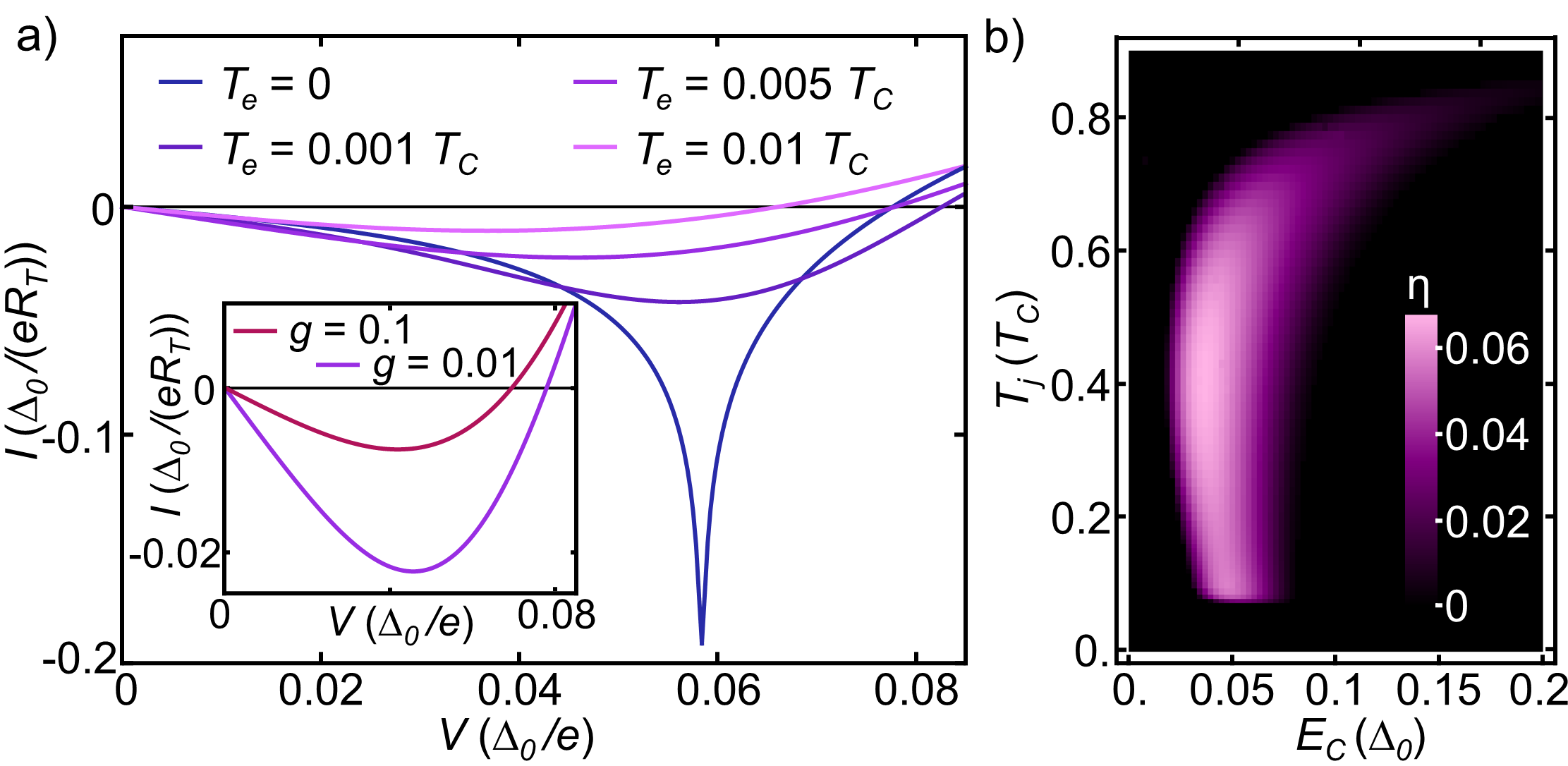}
    \caption{\textbf{Quantum thermoelectricity in realistic conditions} (a) $I$-$V$ characteristics for different values of $T_e$, with $g = 0.01$, $r = 0.9$, $E_C=0.15\Delta_0$ and $T_j = 0.8 T_C$. Inset: comparison between $g = 0.01$ and $g = 0.1$ at $T_e = 0.005 T_C$. (b) Density plot of the  
    efficiency $\eta$ as a function of $E_C$ and $T_j$, computed for $T_e = 0.005T_C$, $g = 0.01$ and $r = 0.9$.} 
    \label{Fig 4}
\end{figure}

We now turn to the role of the environmental temperature, which is directly linked to the \emph{quantum} nature of the thermoelectric effect. To this end, we compute how the temperature of the electromagnetic environment affects quantum thermoelectricity. To model an experimentally realistic setup, we assume a finite value of $g$. In this case, we analytically compute $J(t)$ [see the Methods Section] for the specific $Z_t(\omega)$ and obtain $P(\Delta E)$ by numerical Fourier integration. Figure~\ref{Fig 4}(a) shows the resulting $I$-$V$ characteristic, demonstrating that thermoelectricity is strongly affected by environmental conditions. 

In particular, the peak at $V_P$ broadens and decreases with increasing $T_e$ as a consequence of the progressive reduction of the emission/absorption asymmetry in the environment at higher temperatures. The inset further confirms that increasing $g$ suppresses the thermoelectric signal, reflecting the weakening of DCB effects at lower environmental impedance. This reduction in peak amplitude directly translates into a corresponding decrease in extractable thermoelectric power.
However, excessively increasing the value of environmental resistance also reduces energy exchange between the junction and the reservoir due to the effects of impedance mismatch~\cite{marchegiani_highly_2021}. Therefore, we identify the values of $g\approx 0.01$, corresponding to the resistance of a few M$\Omega$, as the most suitable compromise. The discussed suppression of the peak can be phenomenologically described as a renormalisation of the effective Dynes parameter $\gamma$, which accounts for inelastic processes that limit the quasiparticle lifetimes and smear the DOS. This interpretation is fully consistent with the measured change in the Dynes parameter induced by modifications in the electromagnetic environment~\cite{pekola_environment-assisted_2010}. 

\subsection*{Quantum Thermoelectric Efficiency}
To further clarify the thermoelectric nature of the effect, it is essential to examine its thermodynamic efficiency. In this context, the two reservoirs are the hot junction and the cold electromagnetic environment. A meaningful definition of thermodynamic efficiency applies when the system operates as a power generator, that is, when $IV < 0$, and is given by the ratio of useful electrical power to the total radiative energy transferred from the junction (hot reservoir) to the environment (cold reservoir):
\begin{equation}
\eta = \frac{|I(V)V|}{\mathcal{P}_{e}(V)}. 
\end{equation}
The numerator is obtained directly from the current-voltage characteristic $I(V)$, while the denominator $\mathcal{P}_{e}(V)$ accounts for the photonic heat that flows from the junction to the electromagnetic environment. The latter can be computed by generalising the current expression to keep track of the energy flows instead of the charge~\cite{benenti_fundamental_2017}, weighing the integrand by the energy transferred into the environment $(E - E')$, as can be seen in the Methods Section. This quantity satisfies the energy conservation in the junction, which yields
\begin{equation}
\dot{Q}(V) + \dot{Q}'(V) = -I(V)V + \mathcal{P}_e(V),
\label{conservation}
\end{equation}
where $\dot{Q}$ and $\dot{Q}'$ denote the heat currents flowing out of the two superconducting electrodes generalised within the $P(E)$ theory~\cite{tan_quantum-circuit_2017,marchegiani_nonlinear_2020}.
In Fig.~\ref{Fig 4}(b), we show the behaviour of $\eta$ with respect to $E_C$ and $T_j$, evaluated for the bias that maximises thermopower extraction. As before, a non-linear temperature dependence is observed. However, the efficiency remains appreciable even at extremely low junction temperatures, although the corresponding absolute thermoelectric power is strongly suppressed in this regime. Nevertheless, a minimum value of $T_j\approx 0.1 T_C$ appears to be necessary to activate the process.
The dependence $E_C$ follows a trend similar to Fig.~\ref{Fig 3}(a). Within the framework considered here, the thermodynamic efficiency remains bounded by the Carnot limit~\cite{jarzynski_classical_2004,campisi_nonequilibrium_2015}. However, in the strongly nonlinear regime $T_e \ll T_j$ relevant to this work, the corresponding Carnot efficiency $\eta_C = 1 - T_e/T_j$ is close to unity and therefore not a stringent practical bound.

\subsection*{Resonant Cavity Environment}
The quantum bipolar thermoelectric effect discussed above is not limited to dissipative environments, but can also occur in more structured electromagnetic settings, such as resonant cavities.
As an ideal example, we consider a system coupled to an environment with a single resonant frequency $\omega_{LC}$, modelled in Fig.~\ref{Fig 1}(d) by a pure inductance $L$ in the impedance branch. 
The resulting total impedance has a real part $\text{Re}[Z_t(\omega)] = \frac{\pi}{2C} [\delta(\omega - \omega_{LC}) + \delta(\omega + \omega_{LC})]$, with $\omega_{LC} = 1/\sqrt{LC}$.
The presence of the Dirac delta allows for an analytical evaluation of the $P(E)$ function, as discussed in Ref.~\cite{devoret_effect_1990,ingold_charge_1992} and briefly resumed in the Methods Section. In the zero-temperature limit $T_e \to 0$, this expression reduces to a Poisson series of discrete energy emissions into the cold cavity bath:
\begin{equation}\label{SME_Poisson}
P(E) = \sum_{k=0}^{+\infty} p_k\, \delta(E - k \hbar \omega_{LC}), \qquad
p_k = e^{-\rho} \frac{\rho^k}{k!},
\end{equation}
which describes the probability $p_k$ of emitting $k$ quanta of energy toward the cold resonant cavity. The dimensionless parameter that characterises the strength of the coupling between the environment and the junction is $\rho = E_C / \hbar \omega_{LC}$. This quantity sets both the average number of emitted quanta and the width of the distribution.

\begin{figure}[t!]
    \centering
    \includegraphics[width=\linewidth]{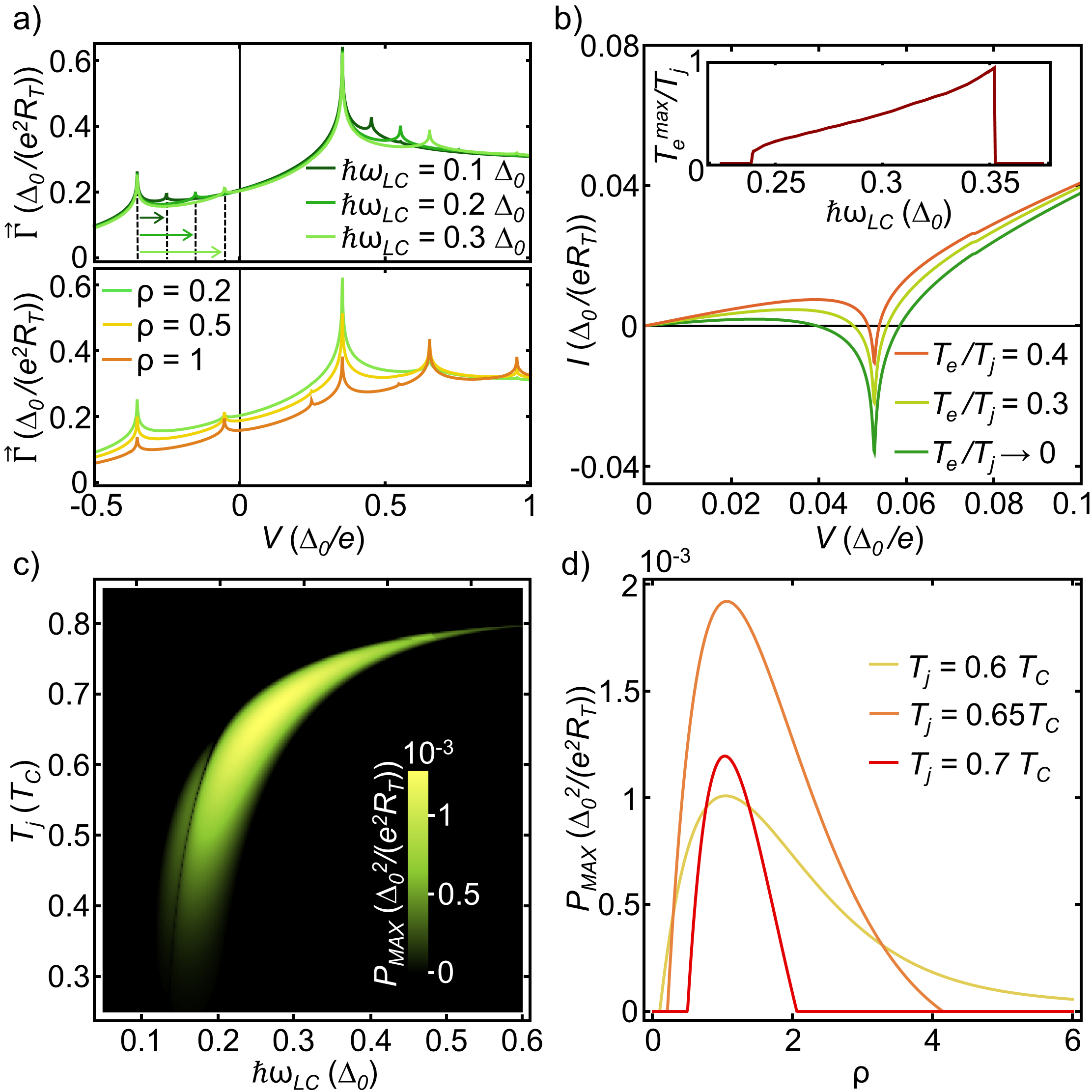}
    \caption{\textbf{Quantum thermoelectricity in a resonant cavity} (a) Top Panel: Forward tunnelling rate $\vec{\Gamma}(\delta\mu)$ for three different environmental resonance frequencies, $\hbar\omega_{LC} = 0.1\Delta_0$ (dark green), $\hbar\omega_{LC} = 0.2\Delta_0$ (green), and $\hbar\omega_{LC} = 0.3\Delta_0$ (light green), with fixed parameters $T_e \to 0$, $T_j = 0.7T_C$, $r = 0.8$, and $\rho = 0.2$. Bottom panel: $\vec{\Gamma}$ as a function of energy for fixed $\hbar\omega_{LC} = 0.3\Delta_0$ and varying coupling strengths: $\rho = 0.2$ (light green), $\rho = 0.5$ (yellow), and $\rho = 1$ (brown), with all other parameters unchanged. (b) $I$-$V$ characteristic for three different environmental temperatures: $T_e \to 0$ (green), $T_e= 0.3 T_j$ (light green), and $T_e = 0.4 T_j$ (red), with $T_j = 0.7T_C$, $r = 0.8$, $\hbar\omega_{LC}=0.3\Delta_0$ and $\rho = 1$. Inset: maximum environmental temperature $T_e^{MAX}$ allowing thermoelectric operation as a function of $\omega_{LC}$ for $T_j=0.7 T_C$ and $r=0.8$. (c) Density plot of the extractable thermoelectric power at the peak ($P_{\mathrm{MAX}}$) as a function of $T_j$ and $\omega_{LC}$, for $r = 0.8$ and $\rho = 0.5$. (d) $P_{\mathrm{MAX}}$ as a function of $\rho$, for three different junction temperatures: $T_j = 0.6T_C$ (yellow), $T_j = 0.65T_C$ (orange), and $T_j = 0.7T_C$ (red), with fixed resonance frequency $\hbar\omega_{LC} = 0.25\Delta_0$ and $r = 0.8$.
}
    \label{Fig 5}
\end{figure}
Figure~\ref{Fig 5}(a) shows the tunnelling rates for an environment described by Eq.~(\ref{SME_Poisson}), for fixed $\rho = 0.2$ and varying $\omega_{LC}$. We observe satellite peaks next to the matching peaks in the forward rate, located at $E=\pm(\Delta - \Delta') + \hbar \omega_{LC}$.  For low values of $\rho$, only the first satellite peak is visible, as multiphoton processes are strongly suppressed, consistent with the Poissonian statistics mentioned.

Unlike microwave-driven setups, where symmetric sidebands develop almost symmetrically around the main peaks~\cite{hijano_microwave-assisted_2023}, the low temperature of the cold bath (quantum regime, $k_B T_e\ll \hbar\omega_{LC}$) inhibits the emission process from the environment, resulting in satellites only at higher energies. In contrast, for photon-assisted processes induced by active driving, an unlimited number of photons is available, and absorption and emission processes are necessarily symmetric. In this classical driving case, bipolar thermoelectricity can be generated \emph{only} if a temperature difference is applied between the two terminals of the junction~\cite{hijano_microwave-assisted_2023}, otherwise the classical photon-assisted current can only be dissipative.
Therefore, for a junction in thermal equilibrium, as considered in this paper, bipolar thermoelectricity can  \emph{only} be generated if the electromagnetic environment operates as a cold quantum bath. This highlights the purely \emph{quantum} nature of the thermoelectric effect discussed here, in sharp contrast with previously reported mechanisms~\cite{germanese_bipolar_2022,hijano_microwave-assisted_2023}.

From the forward rates in the lower panel of Fig.~\ref{Fig 5}(a), we observe that increasing $\rho$ from small values enhances the satellite peaks while progressively suppressing the primary matching peak. For specific parameter values, this can lead to a strong detailed-balance violation, $\vec{\Gamma}(-E)>\vec{\Gamma}(E)$, as previously discussed for the dissipative case, marking the onset of the thermoelectric regime. Higher-order processes that involve multiple quanta may become potentially accessible with sufficiently large values of $\rho$. This results in additional satellite peaks, which may even give rise to other thermoelectric regimes that exhibit increasingly complex behaviour.

Figure~\ref{Fig 5}(b) shows the $I$-$V$ characteristics of this single-mode environment. In particular, we observe that the environmental temperature $T_e$ has a weak effect on the thermoelectric behaviour, indicating greater robustness to thermal broadening than in the purely resistive case discussed in the main text. As shown in the inset, the maximum $T_e$ compatible with a detectable thermoelectric response depends on the resonance frequency: increasing the $\omega_{LC}$ frequency substantially reduces the required thermal difference, until the effect abruptly vanishes.

Figure~\ref{Fig 5}(c) shows the extractable power of the first satellite peak $P_{MAX}$, as a function of $T_j$ and $\hbar \omega_{LC}$ in the regime $k_BT_e\ll \hbar\omega_{LC}$. The behaviour mirrors Fig. 2(a): power increases with $T_j$, peaking near $0.9 T_C'$. Here, $\hbar \omega_{LC}$ sets the position of the satellite and plays a role analogous to $E_C$ in the resistive case, with the optimal range shifting to higher frequencies as the temperature increases. Notably, power vanishes along a curve where a second satellite appears symmetrically opposite the first, causing their contributions to cancel and suppressing the thermoelectric response. 

Figure~\ref{Fig 5}(d) shows that $P_{MAX}$ increases with $\rho$ at fixed $\omega_{LC}$.  The extractable power increases with $\rho$, peaking near $\rho \simeq 1$, then decreases as multiphoton processes become dominant, distributing the bipolar thermoelectric capabilities on multiple channels. At higher $T_j$, the upper tail of the power distribution in $\rho$ shrinks, reflecting the increased likelihood of multi-quanta energy exchange. Compared to the resistive configuration, the maximum extractable power is roughly an order of magnitude lower. 

\section*{Discussion}
We demonstrate the \emph{quantum} bipolar thermoelectric effect emerging from the interaction of an electronic system with a cold electromagnetic environment operating in the quantum regime. The asymmetry between emission and absorption processes disrupts the intrinsic electron-hole symmetry of a superconducting tunnel junction, thereby allowing thermoelectricity to arise spontaneously even when the junction remains locally in thermal equilibrium. Two representative cases were analysed: a resistive (Ohmic) environment, where the charging energy $E_C$ sets the relevant scale, and a single-mode resonant environment, where the resonance frequency $\omega_{LC}$ plays an analogous role. In both cases, the suppression of absorption from the cold bath allows for a strong violation of detailed balance in the quasiparticle rates, enabling a finite thermoelectric output. For realistic parameters of an Nb/AlO$_x$/Nb junction with tunnel resistance $R_T \sim 100~\mathrm{k}\Omega$, gap ratio $r=0.9$ and charging energy $E_C \sim 0.2~\mathrm{meV}$, coupled to a resistive environment with $R \sim 2.5~\mathrm{M}\Omega$ at $T_e \sim 100~\mathrm{mK}$, we estimate a generated power $\sim10~\mathrm{fW}$. The nonlinearity of this effect implies that lowering the temperature of the electromagnetic environment may further enhance the performance. By contrast, the non-linear Seebeck coefficient depends only weakly on the environmental temperature and can reach values up to $\sim 100~\mu\mathrm{V/K}$, depending on the operating regime and system parameters~\cite{germanese_bipolar_2022}. In the parameter range considered for a resistive electromagnetic environment, the radiative power exchanged between the junction and the environment provides the dominant contribution to the energy flow, while the additional Joule heating in the environmental resistor, $I^2R$, is typically smaller, as implied by the low thermodynamic efficiency. 
In practice, optimal device design requires balancing a sufficiently large environmental resistance with efficient electron-phonon thermalization in the resistor, to maintain the electromagnetic environment at low temperature~\cite{giazotto_opportunities_2006}. In contrast, in the superconducting electrodes that form the junction, the electron-phonon coupling is strongly suppressed at low temperatures, allowing the electronic system to be efficiently heated while the phonon bath and the surrounding circuit remain cold, thereby enabling large electronic temperature differences required for thermoelectric operation~\cite{giazotto_opportunities_2006}.\\
Comparable Seebeck values are obtained for realistic resonant environments in the quantum regime. At the same time, the maximum extractable power is typically an order of magnitude smaller than in the resistive case, albeit with greater robustness against environmental temperature variations.
Finally, an important clarification is necessary regarding the single-cavity-mode example. In our calculations, we assume that the cavity mode is perfectly thermalised at a fixed temperature $T_e$. However, this assumption implicitly requires additional degrees of freedom to stabilise the temperature of the cavity mode, even when the hot junction supplies energy. Such coupling to the surrounding environment inevitably reduces the cavity's quality factor, resulting in a finite linewidth of the resonant tone. A fully consistent treatment would therefore require solving the thermalisation problem for the cavity modes and the entire environment self-consistently, including their interactions with the hot junction. This issue has recently been investigated in Ref.~\cite {cailleaux_theory_2025}, which reveals complex and nontrivial dynamics. In other words, our discussion of the resonant cavity completely neglects the thermal feedback from the hot junction, which would likely drive the cavity mode out of equilibrium. Nevertheless, despite this intrinsic limitation, our analysis demonstrates that the quantum bipolar thermoelectric effect is not restricted to a particular form of the electromagnetic environment, but rather represents a generic mechanism. This discussion highlights that the presence of a strong dissipative component in the environment, as in the first example, is likely advantageous for keeping thermal equilibrium, even in the presence of a heat flow from the hot junction, as required by the thermodynamic nature of the discussed quantum bipolar thermoelectric effect.\\ 
These results demonstrate that bipolar quantum thermoelectricity emerges as a genuine quantum effect, enabled by the interplay between superconducting junctions and electromagnetic fluctuations. Beyond the specific examples discussed, the response is inherently sensitive to the spectral and thermal properties of the environment. This suggests that by engineering the electromagnetic environment, the effect can be scaled or tailored to different operational regimes, offering routes towards novel low-temperature thermoelectric devices. At the same time, its dependence on the environment suggests applications in quantum sensing, where superconducting junctions could serve as spectroscopic probes of engineered modes or as detectors of the effective temperature of a quantum bath.

\section*{Methods}
\subsection*{Superconducting properties}
The normalized superconducting density of states (DOS) of electrode $i=S,S'$ is given as
\begin{equation}
N_i(E) = \left| \text{Re} \left[ \frac{E + i\gamma}{\sqrt{(E + i\gamma)^2 - \Delta_i^2(T)}} \right] \right|,
\end{equation}
where $\text{Re}[\cdot]$ denotes the real part of the complex expression and $\gamma$ is the Dynes parameter that accounts for finite quasiparticle lifetime effects~\cite{tinkham_introduction_2004}. In all calculations, we use $\gamma = 10^{-4}\Delta_0$, with $\Delta_0 \equiv \Delta(T=0)$ the zero-temperature superconducting gap. The temperature dependence of the superconducting gap is modeled using the common approximation to the self-consistent BCS solution~\cite{tinkham_introduction_2004},
\begin{equation}
\Delta_i(T) = 1.746\,k_B T_c \,
\tanh\!\left[ 1.74 \sqrt{\frac{T_c}{T} - 1} \right],
\end{equation}
which accurately reproduces the BCS gap over the full temperature range $0<T<T_c$ with a precision of a few \%.

\subsection*{P(E) framework}
In Eq.~(\ref{Intro/rate fin}), we define the forward rate under the influence of the electromagnetic environment in terms of the probability distribution $P(\Delta E)=\frac{1}{2\pi\hbar}\int_{-\infty}^{\infty} dt\, e^{J(t)+ i\Delta E t/\hbar}$. This quantity describes the exchange of an energy $\Delta E$ with the electromagnetic environment. Here, $J(t)=\langle [\tilde{\phi}(t)-\tilde{\phi}(0)]\tilde{\phi}(0)\rangle$ 
is the phase-phase correlator, with $\tilde{\phi}(t)=\phi(t)-(e/\hbar)Vt$ defined in the rotating frame of the bias~\cite{ingold_charge_1992,nazarov_quantum_2009}. 
Its explicit form depends on the total circuit impedance $Z_t(\omega)$ seen from the superconducting tunnel junction. It reads
\begin{align}\label{eq:J_def}
J(t) = &2\int_0^\infty \frac{d\omega}{\omega}\,\frac{\mathrm{Re}[Z_t(\omega)]}{R_K} 
\Bigg[
\coth\!\left(\frac{\hbar\omega}{2k_B T_e}\right)(\cos(\omega t)-1)\nonumber\\&
- i\sin(\omega t)\Bigg],
\end{align}
with $R_K=h/e^2$ the von Klitzing constant and $T_e$ the temperature of the environment.

\subsection*{Energy balance}
The thermodynamic efficiency is determined directly by the balance between heat and electrical power at the junction.  
To express these quantities in a compact form, we introduce the auxiliary function $\Phi_{+}(E,E';V)$
\begin{align} 
\Phi_{+}(E,E';V)= & N_S(E)N_{S'}(E'+eV)\nonumber\\
& f(E)[1-f(E'+eV)], 
\end{align} 
which accounts for forward tunnelling processes. The corresponding backward function $\Phi_{-}(E,E';V)$ is obtained by exchanging the occupation probabilities, $f(E)\to [1-f(E)]$ and viceversa for $f(E'+eV)$.\\
The heat current $\dot{Q}$ flowing out of the S then reads
\begin{align}
\dot{Q}(V)=&\int\frac{dE dE'}{e^2 R_T}
E\Big[\Phi_{+}(E,E';V)P(E-E')-\nonumber\\&-\Phi_{-}(E,E';V)P(E'-E)\Big]. 
\end{align}
The heat current $\dot{Q}'(V)$ flowing out of the S' electrode is obtained in a similar way, with opposite signs for the two terms. It is expressed in terms of $E'$ and accounts for the voltage bias $V$, which necessarily shifts the Fermi energy by a term $eV$. The radiative power transferred to the electromagnetic environment is then obtained by weighting the tunnelling processes with the exchanged energy $\Delta E=E-E'$:
\begin{align}
\mathcal{P}_e(V)=&\int \frac{dE dE'}{e^2 R_T}(E-E')
\Big[\Phi_{+}(E,E';V)P(E-E')-\nonumber\\&-\Phi_{-}(E,E';V)P(E'-E)\Big].
\end{align}
One can easily show that this quantity satisfies the energy conservation at the junction, which is therefore expressed as
\begin{equation}
\dot{Q}(V)+\dot{Q}'(V)=-I(V)V+\mathcal{P}_e(V),
\label{eq:conservation_methods}
\end{equation}
which holds for all bias voltages. This equation shows that the heat losses from the superconducting leads are partially converted into thermoelectric power generation $-I(V) V$ and into a radiative heat flow $\mathcal{P}_e(V)$ from the hot junction to the cold environment.

\subsection*{Ohmic environment}
A purely resistive environment, $Z(\omega)=R$, gives the effective impedance $Z_t(\omega)=(i\omega C+1/R)^{-1}$. Introducing dimensionless conductance $g=R_K/R$, in the high-impedance limit $g\ll1$ and at small $\omega t$, the correlator expands to $J(t)\simeq -\frac{\pi}{R_K C}(t^2/\hbar\beta_e+i t)$, which, after Fourier transform, corresponds to a Gaussian distribution $P(E)=(4\pi E_C k_B T_e)^{-1/2}\exp[-(E-E_C)^2/(4E_C k_B T_e)]$ centred in $E_C$ and with a linewidth proportional to the electromagnetic environment temperature $T_e$. In the quantum limit $T_e\to0$, this reduces to $P(E)=\delta(E-E_C)$.

For finite $g$ and $T_e$, the exact phase correlator can be decomposed into its imaginary and real parts,
\begin{align}
 \mathrm{Im}\,J(t)&=-\frac{\pi}{g}\Big(1-e^{-\omega_R t}\Big),
\end{align}
\begin{align}
 &\mathrm{Re}\,J(t)=-\frac{2\pi}{g}\Bigg[\frac{k_B T_e}{\hbar}t
 -\frac{1-e^{-\omega_R t}}{2}\cot\Big(\frac{\hbar\omega_R}{2k_BT_e}\Big)+\nonumber\\
 &+\sum_{n=1}^{\infty}\frac{2}{\pi n}\frac{1}{1-\Big(2\pi n \frac{k_BT_e}{\hbar\omega_R}\Big)^2}
 \Big(1-e^{- 2\pi n k_BT_et/\hbar}\Big)\Bigg],
\end{align}
with $\cot(x)$ the trigonometric cotangent function and $\omega_R=1/(RC)$. This can be substituted into the definition of $P(E)$ to obtain the numerical evaluation of the response at the junction.

\subsection*{Single-mode environment}
For an electromagnetic environment with a single resonant frequency 
$\omega_{LC}=1/\sqrt{LC}$, it is possible to obtain the analytical result for the function $P(E)$ ~\cite{ingold_charge_1992},
\begin{align}
P(E)=&\exp\!\left[-\rho \coth\!\Big(\tfrac{\beta_e\hbar\omega_{LC}}{2}\Big)\right]
\sum_{k=-\infty}^{+\infty}
I_k\!\left(\frac{\rho}{\sinh(\tfrac{\beta_e\hbar\omega_{LC}}{2})}\right)\nonumber\\
&\times \exp\!\Big(k \tfrac{\beta_e\hbar\omega_{LC}}{2}\Big)\,
\delta(E-k\hbar\omega_{LC}),
\end{align}
where $\rho=E_C/(\hbar\omega_{LC})$ is the dimensionless coupling parameter and $I_k$ 
is the modified Bessel function of the first kind. 
In the quantum limit $T_e\to 0$, this reduces to the Poisson distribution given in Eq.~(\ref{SME_Poisson}).
\section*{Data availability}
The datasets generated during the current study are available in the Zenodo repository at \href{https://doi.org/10.5281/zenodo.16751455}{https://doi.org/10.5281/zenodo.16751455}~\cite{antola_data_2025}.

\section*{Acknowledgments}
A.B. acknowledges the discussions with A. Jordan, B. Bhandari, A. N. Singh, F. Taddei, G. Marchegiani, and the hospitality provided by the Institute for Quantum Studies of Chapman University. The authors acknowledge the PNRR MUR project PE0000023-NQSTI for partial financial support. A.B. acknowledges also the MUR-PRIN2022 Project NEThEQS (Grant No. 2022B9P8LN), the financial support of project ”Thermoelectric effects in solid-state quantum devices based on multiterminal Josephson junctions” of the bilateral agreement CNR/CONICET (Italy/Argentina) 2026-2027, the Royal Society through the International Exchanges between the UK and Italy (Grant No. IEC R2 192166), and the CNR Project QTHERMONANO. The funders played no role in study design, data collection, analysis and interpretation of data, or the writing of this manuscript.

\section*{Competing Interests}
The authors declare no competing interests.

\section*{Author Contributions}
FA performed the numerical simulations and led the development of the project. AB and FG conceived the idea and supervised the work. GDS contributed to the discussion of the results. All authors reviewed and approved the final manuscript.
\bibliography{Bibliography}

\end{document}